\documentclass[sigplan,screen,nonacm]{acmart}

\usepackage{microtype}

\usepackage{hyperref}
\usepackage{xcolor}
\usepackage{amsmath}
\usepackage{mathpartir, xparse}
\usepackage{nameref}

\newcommand{\daml}{Daml}
\newcommand{\damllf}{\daml-LF}
\newcommand{\dalf}{.dalf}

\usepackage{tikz}
\usetikzlibrary{arrows,backgrounds,calc,shapes,positioning}
\usepackage{tikzsymbols}
\usepackage{tikzpeople}

\usepackage{relsize}
\usepackage{listings}
\lstdefinelanguage{Daml}%
  {otherkeywords={=>},%
   morekeywords={abstype,if,then,else,case,class,data,default,deriving,%
      hiding,if,in,infix,infixl,infixr,import,instance,let,module,%
      newtype,of,qualified,type,where,do,AbsoluteSeek,AppendMode,%
      Array,BlockBuffering,BufferMode,Char,Complex,Double,Either,%
      FilePath,Float,Integer,IO,IOError,Ix,LineBuffering,Maybe,%
      Ordering,NoBuffering,ReadMode,ReadWriteMode,ReadS,RelativeSeek,%
      SeekFromEnd,SeekMode,ShowS,StdGen,String,Void,Bounded,Enum,Eq,%
      Eval,ExitCode,exitFailure,exitSuccess,Floating,Fractional,%
      Functor,Handle,HandlePosn,IOMode,Integral,List,Monad,MonadPlus,%
      MonadZero,Num,Numeric,Ord,Random,RandomGen,Ratio,Rational,Read,%
      Real,RealFloat,RealFrac,Show,System,Prelude,EQ,False,GT,Just,%
      Left,LT,Nothing,Right,WriteMode,True,abs,accum,accumArray,%
      accumulate,acos,acosh,all,and,any,ap,appendFile,applyM,%
      approxRational,array,asTypeOf,asin,asinh,assocs,atan,atan2,atanh,%
      bounds,bracket,bracket_,break,catch,catMaybes,ceiling,chr,cis,%
      compare,concat,concatMap,conjugate,const,cos,cosh,curry,cycle,%
      decodeFloat,delete,deleteBy,deleteFirstsBy,denominator,%
      digitToInt,div,divMod,drop,dropWhile,either,elem,elems,elemIndex,%
      elemIndices,encodeFloat,enumFrom,enumFromThen,enumFromThenTo,%
      enumFromTo,error,even,exitFailure,exitWith,exp,exponent,fail,%
      filter,filterM,find,findIndex,findIndices,flip,floatDigits,%
      floatRadix,floatRange,floatToDigits,floor,foldl,foldM,foldl1,%
      foldr,foldr1,fromDouble,fromEnum,fromInt,fromInteger,%
      fromIntegral,fromJust,fromMaybe,fromRat,fromRational,%
      fromRealFrac,fst,gcd,genericLength,genericTake,genericDrop,%
      genericSplitAt,genericIndex,genericReplicate,getArgs,getChar,%
      getContents,getEnv,getLine,getProgName,getStdGen,getStdRandom,%
      group,groupBy,guard,hClose,hFileSize,hFlush,hGetBuffering,%
      hGetChar,hGetContents,hGetLine,hGetPosn,hIsClosed,hIsEOF,hIsOpen,%
      hIsReadable,hIsSeekable,hIsWritable,hLookAhead,hPutChar,hPutStr,%
      hPutStrLn,hPrint,hReady,hSeek,hSetBuffering,hSetPosn,head,%
      hugsIsEOF,hugsHIsEOF,hugsIsSearchErr,hugsIsNameErr,%
      hugsIsWriteErr,id,ioError,imagPart,index,indices,init,inits,%
      inRange,insert,insertBy,interact,intersect,intersectBy,%
      intersperse,intToDigit,ioeGetErrorString,ioeGetFileName,%
      ioeGetHandle,isAlreadyExistsError,isAlreadyInUseError,isAlpha,%
      isAlphaNum,isAscii,isControl,isDenormalized,isDoesNotExistError,%
      isDigit,isEOF,isEOFError,isFullError,isHexDigit,isIEEE,%
      isIllegalOperation,isInfinite,isJust,isLower,isNaN,%
      isNegativeZero,isNothing,isOctDigit,isPermissionError,isPrefixOf,%
      isPrint,isSpace,isSuffixOf,isUpper,isUserError,iterate,ixmap,%
      join,last,lcm,length,lex,lexDigits,lexLitChar,liftM,liftM2,%
      liftM3,liftM4,liftM5,lines,listArray,listToMaybe,log,logBase,%
      lookup,magnitude,makePolar,map,mapAccumL,mapAccumR,mapAndUnzipM,%
      mapM,mapM_,mapMaybe,max,maxBound,maximum,maximumBy,maybe,%
      maybeToList,min,minBound,minimum,minimumBy,mkPolar,mkStdGen,%
      mplus,mod,msum,mzero,negate,next,newStdGen,not,notElem,nub,nubBy,%
      null,numerator,odd,openFile,or,ord,otherwise,partition,phase,pi,%
      polar,pred,print,product,properFraction,putChar,putStr,putStrLn,%
      quot,quotRem,random,randomIO,randomR,randomRIO,randomRs,randoms,%
      rangeSize,read,readDec,readFile,readFloat,readHex,readInt,readIO,%
      readList,readLitChar,readLn,readParen,readOct,readSigned,reads,%
      readsPrec,realPart,realToFrac,recip,rem,repeat,replicate,return,%
      reverse,round,scaleFloat,scanl,scanl1,scanr,scanr1,seq,sequence,%
      sequence_,setStdGen,show,showChar,showEFloat,showFFloat,%
      showFloat,showGFloat,showInt,showList,showLitChar,showParen,%
      showSigned,showString,shows,showsPrec,significand,signum,sin,%
      sinh,snd,sort,sortBy,span,split,splitAt,sqrt,stderr,stdin,stdout,%
      strict,subtract,succ,sum,system,tail,tails,take,takeWhile,tan,%
      tanh,toEnum,toInt,toInteger,toLower,toRational,toUpper,transpose,%
      truncate,try,uncurry,undefined,unfoldr,unionBy,unless,%
      unlines,until,unwords,unzip,unzip3,unzip4,unzip5,unzip6,unzip7,%
      userError,when,words,writeFile,zero,zip,zip3,zip4,zip5,zip6,zip7,%
      zipWith,zipWithM,zipWithM_,zipWith3,zipWith4,zipWith5,zipWith6,%
      zipWith7,%
      template,with,assert,exercise,create,fetch,choice,nonconsuming,%
      controller,signatory,observer,ensure,can,pure},%
   sensitive,%
   morecomment=[l]--,%
   morecomment=[n]{\{-}{-\}},%
   morestring=[b]"%
  }[keywords,comments,strings]%
  
\newcommand{\damlsnippet}[2][]{%
  \lstinputlisting[linerange=#2Begin-#2End, #1]{daml/Iou.daml}%
}

\makeatletter

\DeclareDocumentCommand \inferrule { s O {} m m o }{%
  \IfBooleanTF{#1}%
  {%
    \mpr@inferstar[#2]{#3}{#4}%
  }{%
    \mpr@inferrule[#2]{#3}{#4}%
  }%
  \IfValueT{#5}%
  {%
    \quad
    #5%
    \my@name@inferrule{#5}%
  }%
}
\NewDocumentCommand \my@name@inferrule { m }{%
  \def\@currentlabelname{\ensuremath{#1}}%
}
\makeatother

\usepackage[inline]{enumitem}
\newlist{inlineenum}{enumerate*}{1}
\setlist[inlineenum]{label=(\arabic*)}

\usepackage{subcaption}

\usepackage{comment}

\newcommand{\kindbase}{\star}
\newcommand{\kindarrow}{\mathbin{\rightarrow}}
\newcommand{\tyfun}{\mathbin{\Rightarrow}}
\newcommand{\haskind}{\mathbin{:}}
\newcommand{\hastype}{\mathbin{:}}
\newcommand{\tyapp}{\mathbin{\raisebox{0.2ex}{\text{\footnotesize @}}}}
\newcommand{\wkind}[3]{#1 \mathrel{\vdash} #2 \mathrel{::} #3}
\newcommand{\wty}[3]{#1 \mathrel{\vdash} #2 \mathrel{::} #3}

\newcommand{\lftype}[1]{\mathsf{#1}}
\newcommand{\tyText}{\lftype{Text}}
\newcommand{\tyDecimal}{\lftype{Decimal}}
\newcommand{\tyList}{\lftype{List}}
\newcommand{\lfconst}[1]{\mathsf{#1}}

\newcommand{\tyCid}{\lftype{ContractId}}
\newcommand{\tyUpdate}{\lftype{Update}}
\newcommand{\tyParty}{\lftype{Party}}

\newcommand{\tyIO}{\lftype{IO}}

\begin{document}

\title[\daml: A Smart Contract Language]{\daml: A Smart Contract Language for Securely Automating Real-World Multi-Party Business Workflows}

\author[Bernauer et al.]{
  Alexander Bernauer,
  Sofia Faro,
  R\'emy H\"ammerle,
  Martin Huschenbett,
  Moritz Kiefer,
  Andreas Lochbihler,
  Jussi M\"aki,
  Francesco Mazzoli,
  Simon Meier,
  Neil Mitchell,
  Ratko G. Veprek
}
\affiliation{%
  \institution{Digital Asset}
  \country{Switzerland}
  \vspace*{\baselineskip}
}

\emergencystretch 3em%

\begin{abstract}
  Distributed ledger technologies, also known as blockchains for enterprises, promise to significantly reduce the high cost of automating multi-party business workflows.
  We argue that a programming language for writing such on-ledger logic should satisfy three desiderata:
  \begin{inlineenum}
  \item Provide concepts to capture the legal rules that govern real-world business workflows.
  \item Include simple means for specifying policies for access and authorization. 
  \item Support the composition of simple workflows into complex ones, even when the simple workflows have already been deployed.
  \end{inlineenum}

  We present the open-source smart contract language \daml{} based on Haskell with strict evaluation.
  \daml{} achieves these desiderata by offering novel primitives for representing, accessing, and modifying data on the ledger,
  which are mimicking the primitives of today's legal systems.
  Robust access and authorization policies are specified as part of these primitives,
  and \daml{}'s built-in authorization rules enable delegation, which is key for workflow composability.
  These properties make \daml{} well-suited for orchestrating business workflows across multiple, otherwise heterogeneous parties.

  \daml{} contracts run
  \begin{inlineenum}
  \item on centralized ledgers backed by a database,
  \item on distributed deployments with Byzantine fault tolerant consensus, and
  \item on top of conventional blockchains, as a second layer via an atomic commit protocol.
  \end{inlineenum}
\end{abstract}

\maketitle

\section{Introduction}
\label{sec:introduction}

Automation through distributed applications can increase the efficiency and reduce the cost of conducting business between multiple entities operating in different trust domains.
Distributed ledger technologies (DLT), also known as blockchains for enterprises, promise to significantly reduce the high cost of building such automation, by providing a digital ledger with shared execution logic, as well as atomicity and consistency guarantees for processing the shared data across all participating parties.
The shared logic is typically defined by user-supplied programs called smart contracts, written in specialized programming languages.
To deliver on the promise of easily building such automation, we argue a smart contract language should satisfy three desiderata:

\paragraph{Real-world adequacy}
All business workflows execute in the context of one or more backing legal systems,
whose law code defines the foundational rules for conducting \mbox{business}.
For example, most law code requires contracts to be formed via offer and acceptance \cite{Young2009}, and defines remedies and damages when the contract terms are violated.
The language should provide concepts to capture such rules,
so that they can be seamlessly automated.

In particular, the view ``the code is the law'' is incompatible with the legal systems used to date \cite{Lessig2006}:
Unless society can be convinced to agree that the DLT state represents what people intended to happen,
the state on the ledger cannot be enforced in a court.
While there exist attempts to adjust the legal system accordingly (for a narrow range of use cases),
the alternative solution requires a DLT to support manual intervention to rectify the situation in the event of an unforeseen real-world eventuality \cite{Rose1998SLR}, even if this violates the encoded rules.
The smart contract language should therefore make explicit where, how and by whom intervention can happen during the execution, without requiring apriori knowledge of the exact type of intervention and without relaxing any of the promised security guarantees.

\paragraph{Security}
Smart contracts present a large attack surface for malicious DLT participants.
This is demonstrated by a long and growing list of publicly known smart contract vulnerabilities \cite{dasp_top10,AtzeiBartolettiCimoli2017POST}.
We argue that a secure smart contract language should implement the following requirements:
\begin{itemize}
\item
  Clean language semantics without surprising corner cases helps programmers avoid common security pitfalls.
  For example, unexpected executions such as reentrancy on the Ethereum virtual machine (EVM) are an ongoing security concern.
\item
  Proper authorization checks for changes to the data ensure that misbehaving entities cannot mess with the shared data beyond the intended scope.
  The authorization policy should be specified along with the smart contract code
  so that they can be easily kept in sync.
\item
  The language should be designed for confidentiality such that specific access policies for stored data are easily defined naturally and the smart contract programmer can easily understand and maintain them.
\end{itemize}

\paragraph{Composability}
It should be easy to unilaterally extend the ledger functionality by composing pre-existing workflows into more complex ones.
Composability fosters organic growth of DLT solutions and helps to manage complexity.
We have identified three challenges here:
\begin{itemize}
\item Composition must work even for workflows that have already been deployed and started on a ledger.
\item Delegation of authority is needed so that one entity can execute subworkflows that others have pre-authorized.
\item Validation for a workflow must automatically focus on the relevant subtransaction and ignore the context.
\end{itemize}

In this paper, we present \daml{}, a functional smart contract language designed to satisfy the above desiderata.
\daml{} is open-source and can be obtained from \href{https://www.daml.com}{daml.com}.
It is derived from Haskell and thereby inherits many of its language features such as algebraic datatypes, typeclasses, parametric polymorphism, monads, and higher-order functions.

\daml{} adds novel primitives for representing, accessing, and modifying data on the ledger.
When designing these primitives, we took inspiration from the principles behind the legal systems in force today,
so that we can be confident that common business workflows can be represented in \daml{} naturally.
Moreover, we deliberately do not attempt to introduce new concepts without a corresponding concept in today's legal systems.
For example, trustless bearer tokens and global state invariants have no correspondence in the world of pen-and-paper contracts
and are therefore unsupported.

In \daml{}, the authorization and access policies are specified along with the data and smart contract code as part of the primitives.
The programmer annotates every piece of ledger data with a set of owners and a set of controllers who may change the data according to the smart contract code.
The semantics then ensures that all owners are guaranteed to see the changes.
Moreover, we obtain a clean language semantics by encapsulating the primitives in an $\tyUpdate$ type constructor with monadic operations;
like the $\tyIO$ monad in Haskell, the $\tyUpdate$ monad helps with separating local pure computations from code that depends on and updates the ledger state.

\daml{} is compiled to the core language \damllf{}, which is based on Girard's System F$_\omega$ \cite{Girard1972phd}.
The compiler reuses the Glasgow Haskell Compiler (GHC) frontend to parse, type-check, and desugar the \daml{} code.
Unlike Haskell, \daml{} uses strict call-by-value evaluation that is easier to reason about and to implement;
in a DLT setting, laziness has little benefit as frequent synchronization would trigger evaluation anyway.
\damllf{} code is organized in modules and packages, which can be used from \daml{} with modular compilation.
Code is referenced using content-based addressing, i.e., a cryptographic hash to uniquely identify the referenced code.

The ledger interprets the \damllf{} code using the \daml{} runtime, implemented as a CEK machine with external ledger state.
$\tyUpdate$\ statements result in a transaction tree, a hierarchical description of the ledger state changes.
The tree structure enables workflow composition at the semantics level:
the trees of existing workflows become the subtrees of the combined workflow.
For transaction validation, each constituent workflow can focus on its own subtree and ignore the rest.

\daml{} runs on a variety of ledger implementations while guaranteeing application portability across all of them.
Centralized implementations backed by a relational database systems (e.g., Postgres, Oracle) offer a low barrier to adoption and short development cycles.
For decentralized deployments, Byzantine fault tolerant state machine replication is available for \daml{} on VMware Blockchain~\cite{VMwareBlockchain}
based on the SBFT protocol \cite{GolanGuetaAbrahamGrossmanMalkhiPinkasReiterSeredinschiTamirTomescu2019DSN},
for Hyperledger Fabric \cite{AndroulakiBargerBortnikovCachinChristidisDeCaroEnyeartFerrisLaventmanManevicMuralidharanMurthyNguyenSethiSinghSmithSorniottiStathakopoulouVukolicWeedCocoYellik2018EuroSys},
and for Hyperledger Besu \cite{Besu}, an enterprise version of Ethereum based on IBFT \cite{SaltiniHylandWood2019IBFT},
as well as Corda \cite{Corda}.
At the time of writing, \daml{} smart contracts deployed on such ledgers are used by leading companies in the financial services, healthcare, and supply chain management sectors, processing USD 100b+ in daily transaction volumes.
This indicates that the \daml{} language is fit for purpose.
Moreover, \daml{} provides application portability across these different technologies because the same \daml{} smart contracts run on all those implementations.

Our main contributions are the following:
\begin{itemize}
\item
  The primitives for smart contracts in \daml{} and \damllf{} that abstract the principles found in today's legal systems.
  They enable programmers to easily encode real-world business workflows in smart contract code.
\item
  \damllf{}'s approach to data ownership and its authori\-zation rules for ledger changes,
  which simplify defining write and read access controls for on-ledger data.
\item
  Transaction trees as a model for ledger changes to enable workflow composition after deployment.
\item
  A \daml{} implementation suitable for use in production.
\end{itemize}

The paper is organized as follows:
We introduce the salient features of \daml{} by examples in \autoref{sec:daml:by:example}.
\damllf{}'s syntax, type system, semantics, and authorization rules are given in \autoref{sec:daml-lf}.
We explain implementation aspects of the \daml{} compiler, the \daml{} runtime, and \daml{} ledgers in \autoref{sec:implementation}.
\autoref{sec:related:work} discusses related work;
\autoref{sec:conclusion} concludes.

\section{Daml by Example}
\label{sec:daml:by:example}

\daml{} is an open-source functional smart contract language with Haskell-like syntax.
In this section, we briefly sketch the system model for \daml{} ledgers (\autoref{sec:system:model})
and then present the core concepts and ideas behind the \daml{} language through a simple model of cash IOUs (``I Owe You'') (Sections~\ref{sec:data:modelling}--\ref{sec:composability})
The official documentation with full details is available online at \href{https://docs.daml.com}{docs.daml.com}.

\subsection{System Model}
\label{sec:system:model}

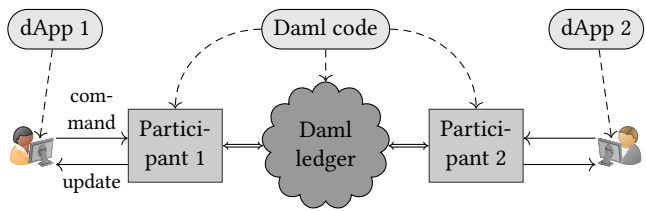
\begin{figure}
  \begin{tikzpicture}[
      on grid, node distance=1.55cm and 2cm,
      message/.style={midway, auto=left, font=\footnotesize},
      font=\small,
      Participant/.style={rectangle,fill=black!20, draw=black!80, inner sep=1ex, align=center},
    ]
    \node (ledger) at (0, 0) [align=center, cloud, draw, fill=black!40, cloud puffs=13] { \daml{}\\ledger };
    \node[Participant] (participant1) [left=of ledger] {Partici-\\pant 1};
    \node[Participant] (participant2) [right=of ledger] {Partici-\\pant 2};

    \node[alice, monitor] (alice) [left=of participant1] {};
    \node[bob, monitor, mirrored] (bob) [right=of participant2] {};

    \begin{scope}[every node/.style={draw, fill=black!10, rounded rectangle, font=\small\vphantom{Ag}}]
      \node (daml) [above=of ledger] { \daml{} code };
      \node (app1) at (current bounding box.west |- daml) [anchor=west] { dApp 1 };
      \node (app2) at (current bounding box.east |- daml) [anchor=east] { dApp 2 };
    \end{scope}

    \draw[->] ([yshift=0.1cm, xshift=1.5ex] alice.east) -- ([yshift=0.1cm] participant1.west) node [message, align=center] { com-\\mand };
    \draw[->] ([yshift=-0.25cm] participant1.west) -- ([yshift=-0.25cm, xshift=1.5ex] alice.east) node [message] { update };

    \draw[->] ([yshift=0.1cm, xshift=-1.5ex] bob.west) -- ([yshift=0.1cm] participant2.east);
    \draw[->] ([yshift=-0.25cm] participant2.east) -- ([yshift=-0.25cm, xshift=-1.5ex] bob.west);

    \begin{scope}[>=implies]
      \draw[<->, double] (participant1) -- (ledger);
      \draw[<->, double] (participant2) -- (ledger);
    \end{scope}

    \begin{scope}[densely dashed]
      \draw[->] (daml)-- (ledger);
      \draw[->] (daml) to[out=180, in=90] (participant1);
      \draw[->] (daml) to[out=0, in=90] (participant2);
      \draw[->, shorten >=0.1cm] (app1) -- (alice.east);
      \draw[->, shorten >=0.1cm] (app2) -- (bob.west);
    \end{scope}
  \end{tikzpicture}
  \caption{System model for \daml{} ledgers}
  \label{fig:system:model}
\end{figure}

\autoref{fig:system:model} shows the system model of a typical \daml{} ledger.
We distinguish between the shared rules and the users' strategies of a distributed application (dApp).
Only the shared rules are implemented in \daml{},
whereas every user decides on their own how to express and maintain their strategy for acting on the ledger.
The strategy part including the dApp frontend is typically implemented in a mainstream language such as Java or TypeScript.
Users interact with the \daml{} ledger by sending \daml{} commands to a ledger client, called a participant.
The participant interprets such a command by running the \daml{} smart contract code to produce a ledger update, which is then forwarded to the \daml{} ledger.
The \daml{} ledger is responsible for enforcing the \daml{} semantics subject to its trust assumptions.
More precisely, the \daml{} ledger
\begin{inlineenum}
  \item validates that ledger updates conform to the \daml{} semantics and
  \item informs the participants affected by a successful ledger update, which in turn notify their users' dApps.
\end{inlineenum}
We discuss \daml{} ledger implementations in \autoref{sec:ledger:integration}.
For now, it suffices to think of a \daml{} ledger as the participant API endpoints that are synchronized.

\subsection{Data Modelling}
\label{sec:data:modelling}

\looseness=1
We start the \daml{} overview with how to model IOUs in \daml{}.
We represent cash amounts using the record type \lstinline!Cash! defined below using the keyword \lstinline!data! with record constructor \lstinline!Cash! and two fields: the \lstinline!currency! of type \lstinline!Text! (i.e., strings) and the \lstinline!amount! of type \lstinline!Decimal!, i.e., a fixed-point number with 28 digits before and 10 digits after the decimal point.

\pagebreak

\damlsnippet{Cash}

An IOU is represented as an instance of a smart contract definition, a \daml{} \lstinline!template!.
Templates are parametrized with the data that is stored in the contract instance, the contract arguments.
For a \lstinline!SimpleIou!, this data consists of the \lstinline!issuer! of the IOU, the current \lstinline!owner! of the IOU, and the \lstinline!cash! that the issuer owes to the owner.
The \lstinline!ensure! clause specifies an invariant on the contract arguments, namely that the cash amount must be positive.

The \daml{} type \lstinline!Party! represents an identity that can act on the ledger, in the sense that they can sign contracts and submit transactions.
How such actions of parties are cryptographically recorded, verified, and evidenced depends on the implementation.
Generally, all \daml{} ledgers must link any transaction to some cryptographic evidence for authorization and non-repudiation purposes.
Linking the evidence to real-world entities is a separate problem, which can be implemented as \daml{} workflows.
This way, we can support the diverse requirements in this field that can not be satisfied with a single approach. 

\damlsnippet{SimpleIou}

The \lstinline!observer! and \lstinline!signatory! clauses specify the notification and authorization rules, resp., for the contract instances of a template:
The observer parties are notified when a contract instance is created or archived.
The non-empty set of signatories specifies the parties whose authority is required to create or archive a contract.
In general, \daml{} ledgers guarantee that a party is notified whenever their authority is used.
For \lstinline!SimpleIou!, the issuer is the signatory and the owner is an observer.
There is no need to specify the issuer as another observer because signatories are notified about creations and archivals anyway, as their authority is used.

\subsection{Encoding Business Logic}

In our example, the IOU owner may transfer the IOU to another party.
Yet, \daml{} contracts use a modified UTXO model and are therefore immutable.
So an IOU transfer archives the old \lstinline!SimpleIou! instance and creates a new \lstinline!SimpleIou! instance for the new owner, atomically within one \daml{} transaction.
Yet, the owner needs the authority of the issuer for both the archival and the creation, because the issuer is the signatory for \lstinline!SimpleIou!s.
To that end, the \lstinline!SimpleIou! template defines a \lstinline!choice SimpleTransfer!.
Choices are code entry points for smart contracts and define the rules prescribing how a contract instance can evolve.
Here, the \lstinline!SimpleTransfer! choice returns a \lstinline!ContractId SimpleIou!, i.e., the identifier of the newly created \lstinline!SimpleIou! contract instance.
The choice takes the new owner as an argument and the keyword \lstinline!controller! says that the owner's authority is needed for exercising this choice, i.e., calling this entry point.
The \lstinline!do! block contains the implementation, namely creating a copy of the current IOU \lstinline!this! with the \lstinline!owner! field updated to \lstinline!newOwner!.
The \lstinline!create! function returns the contract ID the ledger assigns to the newly created instance.

The owner's authority suffices to exercise the \lstinline!SimpleTransfer! choice on the IOU, but creating the transferred IOU requires the issuer's authority (as the issuer is the signatory).
So this demonstrates a form of authority delegation:
When the old IOU was created, the issuer has pre-authorized the consequences of the choices in the template.
Therefore, the issuer's authority is available in the choice implementation and the workflow thus well-authorized.

Every template implicitly defines a parameter-less \lstinline!Archive! choice with all signatories as controllers that simply archives a contract instance.
By default, user-defined choices such as \lstinline!SimpleTransfer! also archive the contract instance they are being exercised on.
Archival means that no further choices can be exercised on the contract.
This \lstinline!Archive! choice ensures that every contract instance can in principle be archived and later deleted from a ledger,
i.e., there are no ``stuck'' \daml{} contracts.
So the signatories can mimic any real-world intervention on the ledger state by jointly archiving obsolete contract instances and appropriately creating new ones.

Yet, this \lstinline!Archive! choice allows the issuer to unilaterally archive a \lstinline!SimpleIou! at any time because the issuer is the only signatory.
Ideally, the owner should consent if an IOU is taken away from them---or if an IOU is given to them, following the propose-accept pattern of contract formation.
We can model this in \daml{} by making the owner also a signatory of the IOU, as shown in the \lstinline!Iou! template.
\lstinline!Iou! differs from \lstinline!SimpleIou! in two points:
\begin{itemize}
\item The \lstinline!owner! is also a signatory.
\item The choice \lstinline!Transfer! has two controllers, the old and the new owner.
\end{itemize}
The joint authority of both controllers is needed to exercise the choice.
This ensures that enough authorization is available for the \lstinline!create! of the new IOU in the choice implementation: the issuer and the new owner are signatories.

\damlsnippet{Iou}

\subsection{The Offer-Accept Pattern}

The \lstinline!Iou! contract instance represents a proper bilateral agreement, equivalent to a legal contract on paper.
To reach this agreement, the issuer can propose the agreement and the future owner can accept the proposal.
We model such a workflow in \daml{} as a separate template \lstinline!IouProposal! where only the issuer is a signatory.
The terms of the IOU are a template parameter of type \lstinline!Iou!, which is the record of the contract arguments of the homonymous template.
The owner of the proposed IOU controls two choices:
\lstinline!Accept! creates the \lstinline!Iou! and returns its contract ID;
and \lstinline!Reject! merely archives the proposal returning the singleton \lstinline!()!.
The implicit \lstinline!Archive! choice allows the issuer to retract its proposal if the issuer neither accepts nor rejects the proposal.

\damlsnippet{IouProposal}

For a \lstinline!Transfer!, the old and new owner to the IOU could run a similar proposal workflow to gather their joint authority for exercising the choice.
For frequent transfers, it is more convenient to setup the workflow once in a role contract template such as \lstinline!IouSendRole!.
Here, the \lstinline!receiver! agrees to being transferred any \lstinline!Iou! from the given \lstinline!sender!, using the \lstinline!Send! choice.
The keyword \lstinline!nonconsuming! means that the \lstinline!Send! choice does not archive the \lstinline!IouSendRole! contract instance
so that it can be used any number of times.
Nonconsuming choices help to reduce contention in UTXO-like models.
If \lstinline!Send! was consuming, the choice body would have to recreate the role contract so that another \lstinline!Iou! can be transferred.
Then, the sender could not send several IOUs concurrently, as they have to wait until they see the fresh contract ID assigned to the recreated role contract.

The choice body in the \lstinline!do! block uses Haskell-style do notation for sequencing the operations on the ledger state.
The \lstinline!fetch! primitive looks up the contract arguments for a given contract ID \lstinline!iouId!;
it fails if the contract does not exist or has already been archived.
The \lstinline!assert! checks that the sender wants to transfer an IOU that they indeed own.
And the \lstinline!exercise! calls the choice \lstinline!Transfer! on the provided contract ID \lstinline!iouId! with the given parameters;
if the contract has already been archived, an \lstinline!exercise! fails like the \lstinline!fetch! primitive.

\damlsnippet{IouSendRole}

\subsection{Composability}
\label{sec:composability}

\daml{} code is organized in modules and packages, and packages can depend on other packages using the \lstinline!import! keyword.
This enables an open business architecture where entities can combine workflows from other entities like building blocks.
For example, a financial institution A can provide the \lstinline!Iou! smart contract with the \lstinline!Transfer! choice in a module \lstinline!BasicAssets! and publish it as a (compiled) \daml{} package \lstinline!Basic!.
Users can already create \lstinline!Iou!s and transfer them.
Later, another entity B can build an atomic swap of two IOUs in another package \lstinline!Swap! by importing \lstinline!Basic!.
As soon as \lstinline!Swap! is deployed, users can atomically swap their existing, unchanged \lstinline!Iou!s.
In particular, A does not have to upgrade its services.
This example demonstrates \daml{}'s composability property.

Suppose that two parties, the initiator and the responder, want to atomically exchange their IOUs, say USD 100 issued by the Fed for CHF 90 backed by the Swiss national bank.
Atomicity is important to reduce the counterparty risk:
without atomicity, if one IOU is transferred before the other, the former owner incurs the risk that the other party defaults or goes out of business before transferring the second IOU; or commits fraud and never transfers the second IOU.

The \lstinline!Transfer! choice already implements the transfer of a single IOU.
So we merely need to compose two such transfers in a single \daml{} transaction.
The \lstinline!TradeProposal! template composes two such transfers, again following the propose-accept pattern.
The initiator specifies the contract ID of its IOU and the \lstinline!Iou! conditions it expects to receive in exchange.
The responder, i.e., the owner of the expected \lstinline!Iou!, can accept the trade proposal and immediately settle the trade with its own \lstinline!Iou! with ID \lstinline!respId!,
where the two \lstinline!exercise! calls to \lstinline!Transfer! aggregate the two transfers into a single transaction.
The keyword \lstinline!pure! at the end defines the result of the body, namely the two new contract IDs.

\pagebreak
\damlsnippet{TradeProposal}

Before accepting the proposal, the responder should check that \lstinline!initId! references an IOU of the initiator that the responder is willing to receive.
Due to \daml{}'s confidentiality model, the responder cannot simply resolve a contract ID and look at the contract arguments,
which may contain confidential business data.
Instead, the initiator can disclose the IOU to the responder by using the nonconsuming choice \lstinline!DiscloseIou!
where the responder is declared as a choice observer.
\daml{} ensures that choice observers are notified whenever the choice is exercised (similar to how observers declared on the template are notified about creation and archival).
Accordingly, the responder observes the \lstinline!fetch!ing of \lstinline!initId! and thereby remembers the mapping from the contract ID to the contract arguments.
This mapping is immutable and thus still valid when the \lstinline!Settle! choice is executed; so there is no need to check that \lstinline!initId! still refers to the same IOU terms.

This concludes the exposition of the \daml{} language features for this paper.
\daml{} provides further primitive types such as strings, dates, timestamps, generic maps, and in particular various numeric types including non-serializable arbitrary-precision numbers for internal calculations.
Other language features not covered in this paper are
algebraic data types, Haskell-style typeclasses, exceptions with try-catch blocks, and contract keys for referencing contracts by value instead of by ID.
They are documented on \href{https://docs.daml.com}{docs.daml.com}.

\section{Daml-LF}
\label{sec:daml-lf}

\daml{} code is compiled into \damllf{} (LF stands for ledger fragment), which is interpreted by \daml{} ledgers.
We now present the core features of \damllf{}, the semantic domain of transaction trees, and the authorization and visibility rules.
The full specification of \damllf{} is available online at
\href{https://github.com/digital-asset/daml/blob/main/daml-lf/spec/daml-lf-1.rst}{github.com/digital-asset/daml/blob/main/daml-lf/spec/daml-lf-1.rst}.
We discuss compilation later in \autoref{sec:daml:compiler}.

\subsection{System F$_\omega$ without Type-Level Lambdas}

\damllf{} is based on System F$_\omega$, an extension of the simply-typed lambda calculus proposed by Girard \cite{Girard1972phd}.
Moreover, it is heavily inspired by GHC Core, the intermediate language used by GHC \cite{sulzmannSystemTypeEquality2007}.
Similar to GHC Core, \damllf{} omits type-level lambdas from System F$_\omega$, and adds user-defined data types.
So type expressions $\tau,\sigma$ are built from type variables $\alpha$,
the function arrow $\tyfun$,
quantification $\forall \alpha \haskind k.\ \tau$,
type application $\tau\;\sigma$,
and primitive type constructors $T$.
Kinds $k$ classify type expressions and rule out ill-formed type expressions, as formalized by the well-kinding judgement $\wkind{\Gamma}{\tau}{k}$.
\begin{trivlist}
\item
  \strut
  \hfill
  \inferrule{
    \alpha \haskind k \in \Gamma
  }{
    \wkind{\Gamma}{\alpha}{k}
  }%
  \qquad
  \inferrule{
  }{
    \wkind{\Gamma}{\tyfun}{\kindbase \kindarrow \kindbase \kindarrow \kindbase}
  }%
  \qquad
  \inferrule{
    \wkind{\Gamma, \alpha \haskind k}{\tau}{k'}
  }{
    \wkind{\Gamma}{\forall \alpha \haskind k.\ \tau}{k'}
  }%
  \hfill
  \strut
  \\*[2\jot]
  \strut
  \hfill
  \inferrule{
    \wkind{\Gamma}{\tau}{k_1 \kindarrow k_2}
    \\
    \wkind{\Gamma}{\sigma}{k_1}
  }{
    \wkind{\Gamma}{\tau\;\sigma}{k_2}
  }%
  \qquad
  \inferrule{
    T_{\mathrm{Prim}}\text{ has kind }k
  }{
    \wkind{\Gamma}{T_{\mathrm{Prim}}}{k}
  }%
  \hfill
  \strut
\end{trivlist}
The primitive type constructors $T_{\mathrm{Prim}}$ include
the singleton type $\lftype{Unit}$,
Booleans $\lftype{Bool}$,
64-bit integers $\lftype{Int64}$,
fixed-point decimals $\tyDecimal$,
strings $\tyText$,
dates $\lftype{Date}$,
and timestamps $\lftype{Timestamp}$, all of kind $\kindbase$,
and the type constructor for lists $\tyList$ of kind $\kindbase \kindarrow \kindbase$.
The primitive type constructors for smart contracts are below presented in \autoref{sec:primitives}.

Terms $t$ are built from variables $x$,
function abstraction $\lambda x \hastype \tau.\ t$, function application $t_1\ t_2$,
type abstraction $\Lambda \alpha \haskind k.\ t$, type application $t \tyapp \tau$, and primitive constants $C$.
All abstractions are annotated by the type or kind so that type checking is decidable.
The typing rules for $\wty{\Gamma}{t}{\tau}$ are standard (\autoref{fig:system:F:typing}).

\begin{figure}
  \strut
  \hfill
  \inferrule{
    x \hastype \tau \in \Gamma
  }{
    \wty{\Gamma}{x}{\tau}
  }%
  \qquad
  \inferrule{
    \wty{\Gamma, x \hastype \tau}{t}{\sigma}
  }{
    \wty{\Gamma}{\lambda x \hastype \tau.\ t}{\tau \tyfun \sigma}
  }%
  \hfill
  \strut
  \\*[2\jot]
  \strut
  \hfill
  \inferrule{
    \wty{\Gamma, \alpha \haskind k}{t}{\tau}
  }{
    \wty{\Gamma}{\Lambda \alpha \haskind k.\ t}{\forall \alpha \haskind.\ \tau}
  }%
  \qquad
  \inferrule{
    \wty{\Gamma}{t_1}{\tau \tyfun \sigma}
    \qquad
    \wty{\Gamma}{t_2}{\tau}
  }{
    \wty{\Gamma}{t_1\;t_2}{\sigma}
  }%
  \hfill
  \strut
  \\*[2\jot]
  \strut
  \hfill
  \inferrule{
    \wty{\Gamma}{\tau}{k}
    \quad\ \
    \wty{\Gamma}{t_1}{\forall \alpha \haskind k.\ \sigma}
  }{
    \wty{\Gamma}{t_1  \tyapp  \tau}{\sigma[\alpha \mapsto \tau]}
  }%
  \hfill
  \strut
\caption{Typing rules for System F$_\omega$}
\label{fig:system:F:typing}
\end{figure}

Well-kinding and well-typing is relative to a set of user-defined \damllf{} types and constants declared in \damllf{} modules.
\damllf{} ensures that every package has a globally unique name (and so do the package's modules):
The package identifier is a cryptographic hash of the \damllf{} code of the package, i.e., a form of content-based addressing.
In this paper, we omit the package identifier and module name for readability and just use the unqualified \daml{} name.
For example, the following \daml{} datatype declaration of binary trees with labelled leaves
\damlsnippet{Tree}
is represented by three \damllf{} type constructors, all of kind $\kindbase \kindarrow \kindbase$:
\begin{itemize}
\item A record type $\lftype{Tree.Leaf}$ with a single field $\lfconst{leaf}$ for the argument of the \lstinline!Leaf! constructor.
\item
  The record type $\lftype{Tree.Node}$ with fields $\lfconst{left}$ and $\lfconst{right}$ for the argument of the \lstinline!Node! constructor.
\item
  The variant type $\lftype{Tree}$ with variants $\lfconst{Leaf}$ and $\lfconst{Node}$.
\end{itemize}

Every record comes with a record constructor, field projections, and field update operations.
Variant types come with constructors and pattern matching support.
The typing rules enforce that these operations are fully applied; if necessary, the \daml{} compiler $\eta$-expands partially applied occurrences.
This convention ensures that all constructors appear fully applied to their arguments, which simplifies the implementation of datatypes.
Again, this is in line with GHC Core.
For example, the typing rules for the record constructor $\lfconst{Tree.Node}$ and the variant constructor $\lfconst{Node}$ are the following:
\begin{center}
  \strut
  \inferrule{
    \wkind{\Gamma}{\tau}{\kindbase}
    \qquad
    \wty{\Gamma}{e_1}{\tau}
    \qquad
    \wty{\Gamma}{e_2}{\tau}
  }{
    \wty{\Gamma}{\lfconst{Tree.Node} \tyapp \tau\ e_1\ e_2}{\lftype{Tree.Node} \tyapp \tau}
  }
  \strut
  \\[2\jot]
  \strut
  \inferrule{
    \wkind{\Gamma}{\tau}{\kindbase}
    \qquad
    \wty{\Gamma}{e}{\lftype{Tree.Node}  \tyapp  \tau}
  }{
    \wty{\Gamma}{\lfconst{Node} \tyapp \tau\ e}{\lftype{Tree} \tyapp \tau}
  }
  \strut
\end{center}

\subsection{Primitives for Smart Contracts}
\label{sec:primitives}

\damllf{} has three type constructors for defining smart contracts:
\begin{itemize}
\item $\tyParty \haskind \kindbase$ for parties, i.e., a subset of strings
\item $\tyCid \haskind \kindbase \kindarrow \kindbase$ for contract identifiers (IDs)
\item $\tyUpdate \haskind \kindbase \kindarrow \kindbase$ for updates of the ledger state
\end{itemize}

Contract IDs are parametrized with a template type like in \daml{};
they uniquely identify contract instances of the template.
Contract IDs are opaque to \daml{} programs so the semantics of \daml{} program is independent of how the \daml{} ledger allocates contract IDs.
They can only be created via $\lfconst{create}$ commands within \damllf{}, but arbitrary values can be passed in as serialized arguments, as explained in \autoref{sec:serializable}.
The primitive comparison operation $\lfconst{\leq}_{\tyCid}$ allows for ordering contract IDs
so that contract IDs can be used in functional data structures such as search trees.

\begin{center}
  \strut
  \inferrule{
    \wkind{\Gamma}{\tau}{\kindbase}
    \\
    \wty{\Gamma}{e_1}{\tyCid\ \tau}
    \\
    \wty{\Gamma}{e_2}{\tyCid\ \tau}
  }{
    \wty{\Gamma}{\lfconst{\leq}_{\tyCid} \tyapp \tau\ e_1\ e_2}{\lftype{Bool}}
  }
  \strut
\end{center}

\newcommand{\lfpure}{\lfconst{pure}}
\newcommand{\lfbind}{\lfconst{bind}}
\newcommand{\lfcreate}{\lfconst{create}}
\newcommand{\lfexercise}{\lfconst{exercise}}
\newcommand{\lffetch}{\lfconst{fetch}}
\newcommand{\lflet}{\lfconst{let}}

The type constructor $\tyUpdate$ of kind $\kindbase \kindarrow \kindbase$ represents the monad of ledger updates.
The monadic operations $\lfpure$ and $\lfbind\ x \hastype \tau \leftarrow e\ \lfconst{in}\ e'$ inject values into the monad and sequence dependent ledger updates.
There are three primitive monad operations: $\lfcreate \tyapp T\ e$, $\lffetch\ \tyapp T\ e$, and $\lfexercise \tyapp T\ \mathit{Ch}\ e_1\ e_2$.
They model (1) creating a contract instance of template $T$ with argument $e$, (2) looking up the contract arguments of $e$, and (3) exercising the choice $\mathit{Ch}$ on $e_1$ with choice argument $e_2$.
As for records and variants, we require those operations to be fully applied.
In the typing rules below, $T$ ranges over templates in the \damllf{} program.
\begin{trivlist}
\item
  \strut
  \hfill
  \inferrule{
    \wkind{\Gamma}{\tau}{\kindbase}
    \qquad
    \wty{\Gamma}{e}{\tau}
  }{
    \wty{\Gamma}{\lfpure \tyapp \tau\ e}{\tyUpdate\ \tau}
  }%
  \hfill
  \strut
  \\[2\jot]
  \strut
  \hfill
  \inferrule{
    \wkind{\Gamma}{\tau_1}{\kindbase}
    \qquad
    \wty{\Gamma}{e_1}{\tyUpdate\ \tau_1}
    \\
    \wty{\Gamma, x \hastype \tau_1}{e_2}{\tyUpdate\ \tau_2}
  }{
    \wty{\Gamma}{\lfbind\ x \hastype \tau_1 \leftarrow e_1\ \lfconst{in}\ e_2}{\tyUpdate\ \tau_2}
  }%
  \hfill
  \strut
  \\[2\jot]
  \strut
  \hfill
  \inferrule{
    \wty{\Gamma}{e}{T}
  }{
    \wty{\Gamma}{\lfcreate \tyapp T\ e}{\tyUpdate\ (\tyCid\ T)}
  }%
  \hfill
  \strut
  \\[2\jot]
  \strut
  \hfill
  \inferrule{
    \wty{\Gamma}{e}{\tyCid\ T}
  }{
    \wty{\Gamma}{\lffetch \tyapp T\ e}{\tyUpdate\ T}
  }
  \hfill
  \strut
  \\[2\jot]
  \strut
  \hfill
  \inferrule{
    \wty{\Gamma}{e_1}{\tyCid\ T}
    \qquad
    \wty{\Gamma}{e_2}{\tau}
    \\
    \text{Choice $\mathit{Ch}$ has argument type $\tau$ and return type $\sigma$}
  }{
    \wty{\Gamma}{\lfexercise \tyapp T\ \mathit{Ch}\ e_1\ e_2}{\tyUpdate\ \sigma}
  }
  \hfill
  \strut
\end{trivlist}

\subsection{Serializable Types}
\label{sec:serializable}

\damllf{} distinguishes between serializable 
and non-seri\-a\-lizable types.
Serializable types are first-order data, free of computation.
In other words, a value of a serializable type is guaranteed to not contain any functions in it.
Contract and choice arguments and choice results must be serializable.
This ensures that all logic involved in a contract is encoded in its definition, and not by its arguments.
While \damllf{} could in theory support higher-order contract arguments, this limitation is desirable for the following reasons:
\begin{itemize}
\item
  Humans can audit \damllf{} contracts and fully understand their implications without making assumptions about their arguments.
  This would be impossible if the arguments contained logic (i.e. functions).
  For example, if functions were allowed to appear in choice arguments,
  the controller could run arbitrary \daml{} code with the signatories' authorization.
\item
  This decoupling between code and data also allows evolving the interpreter and the backing ledger more freely.
  For example, if we want to add optimization passes and just-in-time compilation before interpretation while storing data in contract arguments,
  such optimizations would have to retain enough data to reconstruct the unoptimized \damllf{} representation of the functions so that all parties involved can validate the stored data independent of their optimization settings.
  Similarly, it seems challenging to encode lambdas into arithmetic circuits if we were to support zero-knowledge proofs in the future.
\item
  Isolating the set of types to a small universe of first-order types allows us to store contract data in a variety of databases, from centralized SQL servers to more traditional blockchains; and to establish a mapping with the existing data types for each platform.
  For instance, Daml-LF records can be turned into SQL columns, which in turn allow for efficient querying.
\end{itemize}

Type checking of template declarations enforces that template and choice parameter types are serializable.
The primitive types are serializable and so are lists and contract IDs if their type argument is.
For a \damllf{} record type $R$ to be serializable, the type arguments and every field's type must be serializable.
Similarly for a variant type $V$, the type arguments and each variant's argument type must be serializable.

For example, the \daml{} record \lstinline!Cash! from \autoref{sec:daml:by:example} is serializable because the two fields \lstinline!currency! and \lstinline!amount! have serializable types.
For \lstinline!Tree!, the polymorphic records $\lftype{Tree.Leaf}$ and $\lftype{Tree.Node}$ for the constructor arguments are serializable iff the type argument is serializable, and the same holds for the variant $\lftype{Tree}$.
So $\lftype{Tree}\ \lftype{Int64}$ is serializable.

As non-serializability propagates through type constructors, $\lftype{Tree}\ (\lftype{Int64} \tyfun \lftype{Bool})$ is not serializable either.
Neither are ledger update types $\tyUpdate\ \alpha$ serializable; ledger updates can only be executed against the ledger as described below.

\subsection{Semantics}

\newcommand{\evalexpr}[2]{#1 \mathrel{\Downarrow} #2}
\newcommand{\evalupd}[3]{\langle #1, #2 \rangle \Downarrow #3}
\newcommand{\lfvalue}{\lfconst{Val}}
\newcommand{\lferror}{\lfconst{Err}}

\damllf{} has a call-by-value semantics defined by two big-step evaluation relations for closed terms.
First, expression evaluation $\evalexpr{e}{r}$ produces a result $r$, i.e., an expression value $\lfvalue\ v$ or a fatal error $\lferror\ \mathit{err}$ with an error message $\mathit{err}$.
Expression values $v$ include lambda abstractions and the literal values of built-in datatypes, contract IDs, parties, etc.
Fully applied record and variant constructors form expression values iff all arguments are expression values.
For example,
\begin{equation*}
  \lfconst{Tree.Leaf} \tyapp (\lftype{Int64} \tyfun \lftype{Int64})\ (\lambda x \hastype \lftype{Int64}.\ x + 1)
\end{equation*}
is a value because the lambda abstraction is a value and $\lfconst{Tree.Leaf}$ is fully applied.
Expression evaluation may also return an $\tyUpdate$ expression $u$, which is considered an expression value too.
$\tyUpdate$ expressions come in five forms:
\begin{itemize}
\item $\lfcreate \tyapp T\ v$ for creating a contract instance of template $T$ with value $v$ as contract argument
\item
  $\lfexercise \tyapp T\ \mathit{Ch}\ v_{\mathrm{cid}}\ v_{\mathrm{a}}$ for exercising the choice $\mathit{Ch}$ of template $T$ on contract ID $v_{\mathrm{cid}}$ with choice argument $v_{\mathrm{a}}$
\item $\lffetch \tyapp T\ v$ for looking up the contract arguments of contractd ID $v$
\item $\lfconst{pure} \tyapp \tau\ v$ for a pure value $v$ of type $\tau$
\item $\lfconst{bind}\ x \hastype \sigma \leftarrow v\ \lfconst{in}\ e$ for a (blocked) computation of ledger changes
\end{itemize}

\newcommand{\Active}{\lfconst{Active}}
\newcommand{\Archived}{\lfconst{Archived}}

Second, the $\tyUpdate$ interpretation relation $\evalupd{u}{s}{R}$ interprets the above forms of ledger changes, using expression evaluation for the subterms.
It depends on the current ledger state $s$, which is a map from contract IDs to the following information about the contract instance:
\begin{itemize}
\item The template identifier $T$
\item The contract arguments as an expression value $v$ of a serializable type
\item The signatories $\mathcal{S}$ and observers $\mathcal{O}$
\item The contract state $\mathit{cs}$ ($\Active$ or $\Archived$)
\end{itemize}
A contract ID $\mathit{cid}$ is fresh in $s$ if $s$ is undefined for $\mathit{cid}$.
Update results $R$ are either a value $v$, a transaction $\mathit{tx}$ and an updated state $s'$, or a fatal error $\lferror\ \mathit{err}$ with an error message $\mathit{err}$.

\newcommand{\Create}{\lfconst{Create}}
\newcommand{\Exercise}{\lfconst{Exercise}}
\newcommand{\Fetch}{\lfconst{Fetch}}
\newcommand{\consuming}{\lfconst{consuming}}
\newcommand{\nonconsuming}{\lfconst{nonconsuming}}

Transactions are \daml{}'s hierarchical representation of ledger updates.
A transaction $\mathit{tx}$ is a list of actions, and an action $a$ takes one of the following forms:
\begin{itemize}
\item
  $\Create\ \mathit{cid}\ T\ v\ \mathcal{S}\ \mathcal{O}$ represents the creation of a contract instance of template $T$ with argument $v$ under the contract ID $\mathit{cid}$, with signatories $\mathcal{S}$ and observers $\mathcal{O}$.
\item $\Exercise\ \mathit{cid}\ T.\mathit{Ch}\ v\ K\ \mathcal{C}\ \mathcal{Q}\ \mathcal{S}\ \mathit{tx}$ represents the exercise of choice $\mathit{Ch}$ in template $T$ on contract with ID $\mathit{cid}$ with choice argument $v$.
  The choice kind $K$ is either $\consuming$ or $\nonconsuming$.
  The set $\mathcal{C}$ captures the controllers, $\mathcal{Q}$ the choice observers, and $\mathcal{S}$ the signatories of the exercised contract.
  The $\Exercise$ action contains the result of executing the choice implementation as a subtransaction $\mathit{tx}$, which we refer to as its consequences.
\item $\Fetch\ \mathit{cid}\ T\ \mathcal{S}\ \mathcal{O}$ represents that looking up the contract ID $\mathit{cid}$ found a contract instance of template $T$ with signatories $\mathcal{S}$ and observers $\mathcal{O}$.
\end{itemize}

So actions are roots of trees and a transaction is a forest.
For example, suppose that Alice submits a command to transfer her \lstinline!SimpleIou! issued by the Bank over USD 100, with contract ID $\#1$, to Bob.
This command translates to executing the following \damllf{} expression, where we have omitted all the record clutter introduced by the \daml{} compiler:
\begin{center}
  $\lfexercise \tyapp \lftype{SimpleIou}.\lfconst{SimpleTransfer}\ \#1\ \mathit{Bob}$
\end{center}
Interpreting this expression produces the transaction shown in \autoref{fig:tx:transfer:simple}, with a single root whose consequence is shown as the child.
Controller sets are underlined, signatory sets are in boldface, observer sets are not decorated, empty sets are omitted altogether, and all $\lfexercise$ actions are by default consuming.
The $\lfexercise$ generates the consuming $\Exercise$ node at the root,
and the \lstinline!create! operation in the body of the \lstinline!SimpleTransfer! choice leads to the $\Create$ child node.

\tikzstyle{subaction}=[double, ->, >=implies]

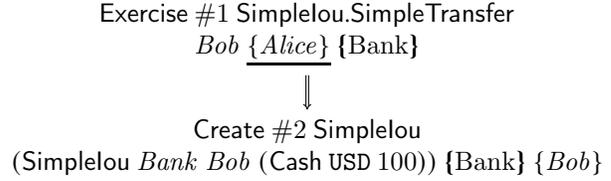
\begin{figure}
  \begin{tikzpicture}
    \node (exerciseTransfer) at (0, 0) [align=center] {
      $\Exercise\ \#1\ \lftype{SimpleIou}.\lfconst{SimpleTransfer}$
      \\
      $\mathit{Bob}\ \underline{\{\mathit{Alice}\}}\ \boldsymbol{\{\mathit{Bank}\}}$
    };

    \node (createIou) at (0, -1.5) [align=center] {
      $\Create\ \#2\ \lftype{SimpleIou}$
      \\
      $(\lfconst{SimpleIou}\ \mathit{Bank}\ \mathit{Bob}\ (\lfconst{Cash}\ \mathtt{USD}\ 100))\ \boldsymbol{\{\mathit{Bank}\}}\ \{\mathit{Bob}\}$
    };

    \draw[subaction] (exerciseTransfer) -- (createIou);
  \end{tikzpicture}

  \caption{Transaction tree for Alice transferring a \lstinline!SimpleIou! of USD 100 to Bob}
  \label{fig:tx:transfer:simple}
\end{figure}

\newcommand{\append}{\mathbin{\mathord+\mathord+}}

$\tyUpdate$ interpretation $\evalupd{u}{s}{R}$ works as follows.
Here, we only discuss the successful case; if any check or sub-evaluation fails, interpretation generates an appropriate error and aborts.

\noindent
For $\lfcreate \tyapp T\ v$:
\begin{enumerate}
\item Find the template definition for $T$.
\item Evaluate the \lstinline!ensure! predicate on $v$ using $\evalexpr{}{}$ and check that the result is $\lfvalue\ \lfconst{True}$.
\item Evaluate the \lstinline!signatory! and \lstinline!observer! functions of $T$ and convert the resulting $\tyParty$ lists into sets $\mathcal{S}$ and $\mathcal{O}$.
\item Pick a contract ID $\mathit{cid}$ that is fresh in $s$.
\item The interpretation result consists of the value $\mathit{cid}$, the transaction $[\Create\ \mathit{cid}\ T\ v\ \mathcal{S}\ \mathcal{O}]$, and the updated state $s[\mathit{cid} \mapsto (T, v, \mathcal{S}, \mathcal{O}, \Active)]$.
\end{enumerate}
For $\lfexercise \tyapp T\ \mathit{Ch}\ v_{\mathrm{cid}}\ v_{\mathrm{arg}}$:
\begin{enumerate}
\item
  Look up the contract data in the state $s$, say $s(v_{\mathrm{cid}}) = (T', v_{\mathrm{c}}, \mathcal{S}, \mathcal{O}, \mathit{cs})$,
  and check that $T' = T$ and $\mathit{cs} = \Active$.
  Set $K = \consuming$ if $\mathit{Ch}$ is consuming and $K = \nonconsuming$ otherwise.
\item
  Evaluate the \lstinline!controller! and \lstinline!observer! expression over the choice $\mathit{Ch}$ on the contract argument $v_{\mathrm{c}}$ and the choice argument $v_{\mathrm{arg}}$ and convert the resulting $\tyParty$ lists into sets $\mathcal{C}$ and $\mathcal{Q}$.
\item
  Substitute $v_{\mathrm{c}}$ and $v_{\mathrm{arg}}$ for the formal template and choice parameters in the choice body expression and evaluate it to an update expression $u'$.
\item
  Interpret $u'$ in state $s'$ where
  $s' = s[v_{\mathrm{cid}} \mapsto (T, v_{\mathrm{c}}, \mathcal{S}, \linebreak  \mathcal{O}, \Archived)]$ if $\mathit{Ch}$ is $\consuming$ and $s' = s$ otherwise.
  The result of interpreting $u'$ consists of a value $v_r$, a transaction $\mathit{tx}$, and a new state $s''$.
\item
  The result of interpreting the $\Exercise$ then consists of $v_r$, the transaction $\Exercise\ v_{\mathrm{cid}}\ T.\mathit{Ch}\ v_{\mathrm{arg}}\ K\ \mathcal{C}\ \mathcal{Q}\ \mathcal{S}\ \mathit{tx}$, and the state $s''$.
\end{enumerate}
For $\lffetch \tyapp T\ v$:
\begin{enumerate}
\item
  Look up the contract data in $s$, say $s(v) = (T', v_{\mathrm{c}}, \mathcal{S}, \mathcal{O}\, \mathit{cs})$,
  and check that $T = T'$ and $\mathit{cs} = \Active$.
\item
  The result consists of the value $v_{\mathrm{c}}$, the transaction $[\Fetch\ v_{\mathrm{cid}}\ T\ \mathcal{S}\ \mathcal{O}]$, and the unchanged state $s$.
\end{enumerate}
For $\lfpure \tyapp \tau\ v$, the result is $(v, [], s)$.
Sequencing $\lfbind\ x \hastype \sigma \leftarrow v\ \lfconst{in}\ e$ composes the resulting transactions:
\begin{enumerate}
\item
  Interpret $v$ in $s$, say $\evalupd{v}{s}{(v_1, \mathit{tx}_1, s_1)}$.
\item
  Substitute $v_1$ for $x$ in $e$ and evaluate it, $\evalexpr{e[x \mapsto v_1]}{u}$.
\item
  Interpret $u$ in $s_1$, say $\evalupd{u}{s_1}{(v_2, \mathit{tx}_2, s_2)}$.
\item
  The result then is $(v'', \mathit{tx}_1 \append \mathit{tx}_2, s'')$,
  where $\append$ concatenates two lists.
\end{enumerate}

As can be seen, the ledger state of a contract ID is updated at most twice:
Once when the contract ID is picked for a new contract instance and once when a consuming choice is exercised on it.
Such immutable contract instances are reminiscent of Bitcoin's UTXO model of transaction outputs,
with the difference that \daml{} contract instances can store arbitrary data and not just an amount and a validation function.
We compare \daml{}'s execution model in more detail to related work in \autoref{sec:related:work}.

\subsection{Turing Completeness}

\damllf{} expressions do not contain a recursive binder such as Haskell's \lstinline!let! or OCaml's \lstinline!let rec!.
Instead, a \damllf{} module can contain an arbitrary number of top-level \emph{value definitions}, which bind a name to a \damllf{} expression, as opposed to defining a contract.
Value definitions can be mutually recursive and therefore endow \damllf{} with unrestricted recursion and make it Turing complete.

Moreover, \damllf{} admits non-monotonic datatypes, which can also be used to define fixpoint combinators without explicit recursion, as shown by the $Z$ fixpoint combinator \lstinline!fix!:

\damlsnippet{Fixpoint}

Making \damllf{} total would not bear any practical ad\-van\-tages\mbox{---}in contrast to type-theory based theorem provers, where soundness relies on a total, non-turing complete language.
What matters for DLT applications is the time it takes to compute a function; and
total languages do not exclude writing functions that take thousands of years to compute.
Instead, ledger implementations may assign a cost to each \damllf{} evaluation step and use this cost model for resource management, similar to Ethereum's gas model \cite{Wood2014ETH}.

\subsection{Authorization Rules}

\newcommand{\wauth}[2]{#1 \vdash_{\mathrm{a}} #2}

In \autoref{sec:daml:by:example}, we sketched \daml{}'s authorization rules with signatories and controllers.
Yet, authorization is not checked by the evaluation semantics from the previous section.
Instead, we define the authorization rules on transactions.
We write $\wauth{\mathcal{A}}{a}$ and $\wauth{\mathcal{A}}{\mathit{tx}}$ if the action $a$ or the transaction $\mathit{tx}$ is well-authorized in the authorization context $\mathcal{A}$, where $\mathcal{A}$ a set of parties.
When a party $A$ submits a \daml{} command to a ledger, the authorization context for the resulting transaction as a whole consists only of $\{A\}$.
However, the authorization context can grow and change for sub(trans)actions of $\Exercise$ actions.

The formal authorization rules are given below.
A $\Create$ action is well-authorized in the context $\mathcal{A}$ iff $\mathcal{A}$ contains all signatories.
For a $\Fetch$, $\mathcal{A}$ must contain at least one signatory or observer of the fetched contract.
The rule for $\Exercise$ is what enables delegation:
First, $\mathcal{A}$ must contain all controllers, as expected.
Second, well-authorization of the consequences is checked in the context of the signatories and controllers.
This encodes that a signatory has pre-authorized the consequences of the choices when the contract instance was created.
Delegation is not transitive, though, as the authorization context $\mathcal{A}$ of the $\Exercise$ action does not propagate to the consequences.
Finally, a transaction is well-authorized if all its (top-level) actions are.
\begin{mathpar}
  \inferrule{
    \mathcal{S} \subseteq \mathcal{A}
  }{
    \wauth{\mathcal{A}}{\Create\ \mathit{cid}\ T\ v\ \mathcal{S}\ \mathcal{O}}
  }

  \inferrule{
    \mathcal{A} \cap (\mathcal{S} \cup \mathcal{O}) \neq \{\}
  }{
    \wauth{\mathcal{A}}{\Fetch\ \mathit{cid}\ T\ \mathcal{S}\ \mathcal{O}}
  }

  \inferrule{
    \mathcal{C} \subseteq \mathcal{A}
    \\
    \wauth{\mathcal{S} \cup \mathcal{C}}{\mathit{tx}}
  }{
    \wauth{\mathcal{A}}{\Exercise\ \mathit{cid}\ T.\mathit{Ch}\ v\ K\ \mathcal{C}\ \mathcal{Q}\ \mathcal{S}\ \mathit{tx}}
  }

  \inferrule{
    \forall a \in \mathit{tx}.\ \wauth{\mathcal{A}}{a}
  }{
    \wauth{\mathcal{A}}{\mathit{tx}}
  }%
\end{mathpar}

\begin{figure}
  \begin{tikzpicture}[
      node distance=0.5cm and 0.6cm,
    ]
    \node (exerciseSettle) at (0, 0) [align=center] {
      $\Exercise\ \#3$
      \\
      $\lftype{TradeProposal}.\lfconst{Settle}\ \#4$
      \\
      $\underline{\{C\}}\ \boldsymbol{\{A\}}$
    };

    \node (fetchIou) [below left=of exerciseSettle, align=center] {
      $\Fetch\ \#4$
      \\
      $\lftype{Iou}\ \boldsymbol{\{C, D\}}$
    };

    \node (exerciseTransferR) [below=of exerciseSettle, align=center] {
      $\Exercise\ \#4$
      \\
      $\lftype{Iou}.\lfconst{Transfer}\ A$
      \\
      $\underline{\{A, C\}}\ \boldsymbol{\{C, D\}}$
    };

    \node (createIouI) [below=of exerciseTransferR, align=center] {
      $\Create\ \#5$
      \\
      $\lftype{Iou}\ \ldots\ \boldsymbol{\{A, D\}}$
    };

    \node (exerciseTransferI) [below right=of exerciseSettle, align=center] {
      $\Exercise\ \#2$
      \\
      $\lftype{Iou}.\lfconst{Transfer}\ C$
      \\
      $\underline{\{A, C\}}\ \boldsymbol{\{A, B\}}$
    };

    \node (createIouR) [below=of exerciseTransferI, align=center] {
      $\Create\ \#6$
      \\
      $\lftype{Iou}\ \ldots\ \boldsymbol{\{B, C\}}$
    };

    \draw[subaction] (exerciseSettle) -- (fetchIou);
    \draw[subaction] (exerciseSettle) -- (exerciseTransferR);
    \draw[subaction] (exerciseSettle) -- (exerciseTransferI);
    \draw[subaction] (exerciseTransferR) -- (createIouI);
    \draw[subaction] (exerciseTransferI) -- (createIouR);

  \end{tikzpicture}
  \caption{Transaction for $C$ exercising the \lstinline!Settle! choice on a \lstinline!TradeProposal!\ by $A$ (contract ID $\#3$) for swapping an \lstinline!Iou! issued by $B$ ($\#2$) against one issued by $D$ ($\#4$)}
  \label{fig:transaction:settle}
\end{figure}
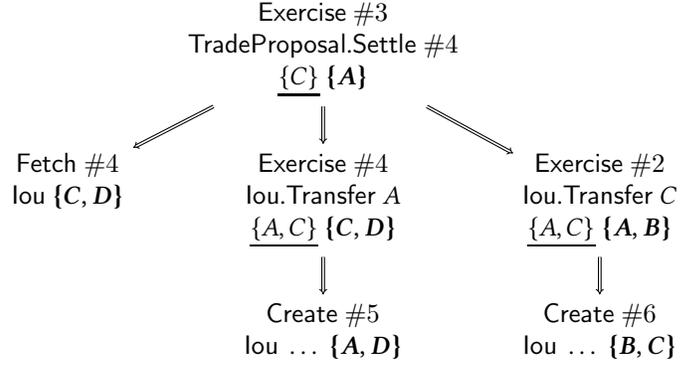

Delegation not being transitive is a deliberate design choice in \daml{}.
It simplifies reasoning about well-authorization of \daml{} programs
because the authorization context for a choice implementation is determined solely by the template definition and thus independent of the calling context of the choice.
On the one hand, this helps \daml{} programmers when they wonder whether sufficient authority is available in a choice implementation.
On the other hand, it is also a security feature important for modularly composing workflows:
When workflow building blocks (i.e., choices) are composed into more complex ones as shown in the \lstinline!Settle! choice,
the authorization safety analysis for the building blocks remains unaffected.
If the consequence of an $\Exercise$ action were instead checked in the augmented context $\mathcal{A} \cup \mathcal{S} \cup \mathcal{C}$,
then the additional parties in $\mathcal{A}$ could make a subaction well-authorized that in isolation would not be.

If transitive delegation is needed, the required signatories must be declared as explicit controllers of the \lstinline!exercise!d choices in the choice body.
This pattern can be seen in the \lstinline!Settle! choice where the \lstinline!initiator!'s authority is passed through the \lstinline!exercise! on \lstinline!respId! via the \lstinline!newOwner! controller of the \lstinline!Transfer! choice to the creation of the new \lstinline!Iou!.
\autoref{fig:transaction:settle} shows the transaction where $C$ exercises the \lstinline!Settle! choice on a \lstinline!TradeProposal! (contract ID $\#3$) where $A$ offers $C$ to trade the \lstinline!Iou! with ID $\#2$ issued by $B$ for an \lstinline!Iou! issued by $D$ ($\#4$).
This transaction is well-authorized in the context $\{C\}$.
Crucially, $A$'s authority from being a signatory on the \lstinline!TradeProposal! contract $\#3$ joins $C$'s authority from being the controller of the $\Exercise$ in the controllers $\{A, C\}$ of the $\Exercise\ \#4$ so that $A$'s authority is available when creating $\#5$.

\section{Implementation}
\label{sec:implementation}

We now describe how we have implemented the \daml{} compiler from \daml{} to \damllf{} (\autoref{sec:daml:compiler}) and the \damllf{} runtime (\autoref{sec:daml:engine}), and sketch how the \damllf{} runtime has been integrated into different ledgers (\autoref{sec:ledger:integration}).

\subsection{\daml{} Compiler}
\label{sec:daml:compiler}

The \daml{} compiler compiles the \daml{} modules in a \daml{} package into a \dalf{} file, a Protobuf encoding of the generated \damllf{} code.
The compiler uses the Glasgow Haskell Compiler (GHC) frontend to parse, type-check, and desugar the \daml{} code into GHC's intermediate language GHC Core, a variant of System FC \cite{WeirichHsuEisenberg}.

Initially, we implemented a \daml{} compiler from scratch, designing the language, writing parsers, type checkers and a desugarer.
However, adding all the syntactic conveniences and language features was a daunting challenge.
With every new feature, new bugs were introduced, especially around parsing and type checking.
So we changed track and decided to reuse the frontend of GHC, a well-established compiler that has benefited from many decades of development.
As a result, \daml{} now provides many of Haskell's language features and syntactic conveniences that make developers productive,
e.g., typeclasses and pattern matching with guards.

At the same time, we recognised there were weaknesses in Haskell that were very apparent when developing \daml{}.
We thus set out to improve these aspects both for \daml{} users and for Haskell users.
In particular, the Haskell record system has long been recognised as somewhat weak compared to other languages, so we started the work on the \lstinline!RecordDotSyntax! proposal, along with many other members of the open source Haskell community.
It is now available in the most recent release of GHC (version 9.2) \cite{GHCproposal0282}.
We also developed a \daml{}/Haskell IDE called ghcide.
This project was later merged with another IDE project to produce the Haskell Language Server \cite{haskell_language_server}, which has become the standard Haskell IDE.

\begin{figure}
  \begin{tikzpicture}[
      on grid,
      node distance=1.5cm and 1.45cm,
      stage/.style={draw, minimum width=2.4cm, font=\vphantom{Ag}, fill=black!15},
      shared stage/.style={stage, minimum width=5.3cm},
      data/.style={font=\footnotesize\vphantom{Ag}, text=black, inner ysep=0.1ex, align=left},
      runtime/.style={fill=black!5, densely dashed, align=center},
      unused/.style={opacity=0.5},
    ]

    \node (middle) at (0, 0) {};
    \node[shared stage] (lexer) [below=0.75cm of middle] { lexer + parser };
    \node[stage] (preprocessor) [below right=of lexer] { preprocessor };
    \node[shared stage] (renamer) [below left=of preprocessor] { renamer + type checker };
    \node[shared stage] (desugarer) [below=of renamer] { desugarer };

    \node[stage, unused] (simplifier) [below left=of desugarer] { simplifier };
    \node[stage, unused] (stg) [below=of simplifier] { STG converter };
    \node[stage, unused] (code-gen) [below=of stg] { code generator };

    \node[stage] (converter) [below right=of desugarer] { LF converter };
    \node[stage] (postprocessor) [below=of converter] { postprocessor };
    \node[stage] (encoder) [below=of postprocessor] { encoder };

    \node[data, unused] (hs-input)  [left=of middle, anchor=west] { Haskell\\ (Text) };
    \node[data] (daml-input) [right=of middle, anchor=west] { \daml{} \\ (Text) };
    \node[data, unused] (assembly) [below=0.75cm of code-gen, anchor=west] { assembly \\ (machine code) };
    \node[data] (dalf) [below=0.75cm of encoder, anchor=west] { \dalf{} \\ (Protobuf) };

    \draw[->] (hs-input.north west) -- (lexer.north -| hs-input.west);
    \draw[->] (daml-input.north west) -- (lexer.north -| daml-input.west);
    \draw[->] (lexer.south -| preprocessor) -- (preprocessor) node[data, midway, anchor=west] { parser tree \\ (AST) };
    \draw[->] (preprocessor) -- (renamer.north -| preprocessor) node[data, midway, anchor=west] { parser tree \\ (AST) };
    \draw[->] (lexer.south -| hs-input.west) -- (renamer.north -| hs-input.west) node[data, midway, anchor=west] { parser tree\\ (AST) };
    \draw[->] (renamer) -- (desugarer) node[data, midway, anchor=west] { typechecker tree \\ (AST) };

    \draw[->, unused] (desugarer.south -| simplifier) -- (simplifier) node[data, midway, anchor=west] { GHC Core \\ (AST) };
    \draw[->, unused] (simplifier) -- (stg) node[data, midway, anchor=west] { GHC Core \\ (AST) };
    \draw[->, unused] (stg) -- (code-gen) node[data, midway, anchor=west] { STG \\ (AST) };
    \draw[->, unused] (code-gen) -- (assembly.south west);

    \draw[->] (desugarer.south -| converter) -- (converter) node[data, midway, anchor=west] { GHC Core \\ (AST) };
    \draw[->] (converter) -- (postprocessor) node[data, midway, anchor=west] { \damllf{} \\ (AST) };
    \draw[->] (postprocessor) -- (encoder) node[data, midway, anchor=west] { \damllf{} \\ (AST) };
    \draw[->] (encoder) -- (dalf.south west);

    \node (right-middle) [right=of postprocessor.south] {};
    \node[stage, runtime] (type-checker) [right=of right-middle, anchor=south] { \damllf{} \\ type checker };
    \node[stage, runtime] (decoder) at (encoder -| type-checker) { decoder };
    \node[stage, runtime] (speedy) at (converter.north -| type-checker) [anchor=south] { compiler };
    \node[stage, runtime] (runtime) [above=of speedy] [anchor=south] { \damllf{} \\ interpreter };
    
    \node[cylinder, shape border rotate=90, shape aspect=0.5, draw=white, thick, fill=black, text=white, minimum width=1.75cm] (ledger)
    at ([yshift=2.3cm] runtime.north east) [anchor=east] {
      ledger
    };

    \node (submission) at (daml-input.north -| runtime.150) {};
    \node[data] (command) at (submission) [anchor=north west, align=left] { command \\ + parties };
    
    \begin{scope}[densely dashed]
      \draw[->] (dalf.south -| decoder) -- (decoder) node [pos=0, data, anchor=south west] { \dalf{} \\ (Protobuf) };
      \draw[->] (decoder) -- (type-checker) node[data, midway, anchor=west] { \damllf{} \\ (AST) };
      \draw[->] (type-checker) -- (speedy) node[data, midway, anchor=west] { \damllf{} \\ (AST) };
      \draw[->] (speedy) -- (runtime) node[data, midway, anchor=west] { ANF \\ (AST) };
    \end{scope}
    \draw[->] (command.north west) -- (command.north west |- runtime.north);
    \draw[<-] (ledger.230 |- runtime.north) -- (ledger.230) node [data, align=center, rotate=90, anchor=center, midway] { contract \\ instances };
    \draw[->] (ledger.310 |- runtime.north) -- (ledger.310) node [data, anchor=north, rotate=90, midway] { transaction };

    \draw[->, thick] (postprocessor.east) -- (postprocessor -| type-checker.west);

  \end{tikzpicture}
  \caption{Pipelines for GHC (left), the \daml{} compiler (middle), and the \damllf{} runtime (right)}
  \label{fig:compiler:pipeline}
\end{figure}
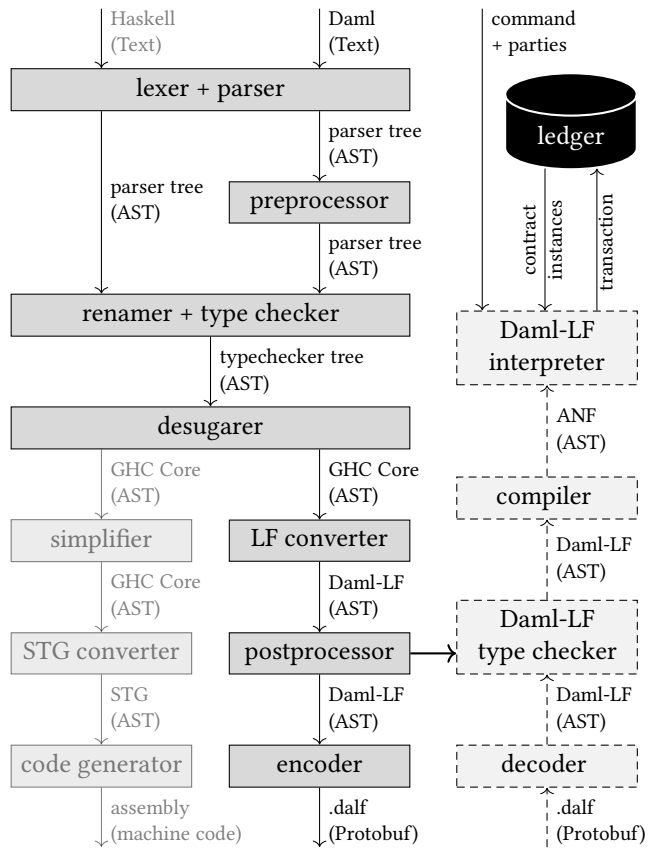

\autoref{fig:compiler:pipeline} shows, from left to right, the pipelines of GHC, the \daml{} compiler, and the \damllf{} runtime, which we discuss later in \autoref{sec:daml:engine}.
Pipeline stages are shown as rectangles and the arrows are labelled by the data that flows from one stage to the next.
The wide rectangles are shared between GHC and the \daml{} compiler, i.e., the lexer, parser, renamer, type checker, and the desugarer.
The renamer, type checker, and desugarer are exactly the same.

For \daml{}'s \lstinline!template! and \lstinline!choice! syntax, we have extended the GHC lexer and parser with appropriate parse rules.
The modified parser desugars templates and choices into Haskell typeclasses
so that the later stages can remain unchanged.
Each template operation has their own Haskell typeclass
so that upgrading is easy when we add more operations.
\begin{lstlisting}[language=Haskell, gobble=2]
  class HasEnsure t where ensure :: t -> Bool
  class HasObserver t where observer :: t -> [Party]
  class HasSignatory t where signatory :: t -> [Party]
  class HasCreate t where
    create :: t -> Update (ContractId t)
  class HasFetch t where fetch :: ContractId t -> Update t
  class HasExercise t c r | t c -> r where
    exercise :: ContractId t -> c -> Update r
\end{lstlisting}
The \lstinline!HasExercise! typeclass is parametrized by the template \lstinline!t! and the choice argument \lstinline!c! and the choice result \lstinline!r!, where the combination of \lstinline!t! and \lstinline!c! uniquely determines \lstinline!r!.
This works because every choice generates a separate record type for the choice arguments.
The parser also generates appropriate instance declarations for all declared templates and choices.

For example, the \lstinline!SimpleIou! template becomes the Haskell datatype for the template arguments with typeclass instances for \lstinline!HasEnsure!, \lstinline!HasObserver!, etc.
The primitive \lstinline!Update! operations retained as type-level strings until the LF conversion phase replaces them with the appropriate \damllf{} primitives.
The \lstinline!SimpleTransfer! choice similarly yields a datatype for the choice argument with an \lstinline!HasExercise! instance.

\begin{lstlisting}[language=Haskell,gobble=2]
  data SimpleIou = SimpleIou with
    issuer : Party
    owner : Party
    cash : Cash

  instance HasEnsure SimpleIou where
    ensure this = (cash.amount this) > 0.0
  instance HasObserver SimpleIou where
    observer this = [owner this]
  instance HasSignatory SimpleIou where
    signatory this = [issuer this]
  instance HasCreate SimpleIou where
    create = primitive @ "UCreate"
  instance HasFetch SimpleIou where
    fetch = primitive @ "UFetch"

  data SimpleTransfer = SimpleTransfer with
    newOwner : Party

  instance HasExercise
      SimpleIou SimpleTransfer (ContractId SimpleIou)
    where exercise = primitive @ "UExercise"
\end{lstlisting}

The new preprocessor provides early warnings and errors of Haskell features that are not available in \daml{}, e.g., unsupported language extensions like GADTs.
It also injects \daml{}-specific imports into the modules so that the later stages can handle the generated template code.

Next, the abstract syntax tree (AST) is processed by the renamer, type checker, and desugarer.
The desugarer in particular replaces typeclasses with dictionaries \cite{HallHammondPeytonJonesWadler1994ESOP}.
Thereafter, the AST is in GHC Core and the \daml{} compiler departs from GHC.
The GHC simplifier, which optimizes GHC Core, is unfortunately unusable as Haskell's lazy evaluation is baked deeply into the optimization rules.
Instead, the LF converter transforms GHC Core into \damllf{}.
Thereby, it converts these typeclass instances back into \damllf{} templates and choices and it also replaces the \lstinline!primitive! placeholder with the appropriate primitive operations in \damllf{}.
The conversion fails if GHC Core uses features from System FC that are not present in \damllf{}, e.g., GADTs.

The rest of the pipeline is mostly standard:
The postprocessor first cleans up the generated \damllf{} code and applies \damllf{}-specific optimizations.
Then, it typechecks the resulting \damllf{} code,
as an extra safe-guard against bugs in the compiler stages.
Finally, the encoder serializes the code as a binary Protobuf message, which is stored as a \dalf{} file.

We have made sure that builds with the \daml{} compiler are repeatable when run in single-threaded mode.
That is, if the same \daml{} input program is compiled with the same version of the \daml{} compiler,
then the same \dalf{} file will be generated.
This is key for content-based addressing in \damllf{},
as every entity can simply recompile a \daml{} package if they do not want to trust the source of a shared \dalf{} file.

\subsection{\damllf{} Runtime}
\label{sec:daml:engine}

The \damllf{} interpreter sits at the heart of the \damllf{} runtime.
As illustrated in \autoref{fig:compiler:pipeline} on the right,
its input are commands of the form \lstinline!create T v! or \lstinline!exercise cid Ch v! for creating a contract of template \lstinline!T! or exercising a choice \lstinline!Ch! on contract ID \lstinline!cid! with argument \lstinline!v!.
It also receives the list $\mathcal{A}$ of parties submitting the commands.
During interpretation, it queries the \daml{} ledger (or a cache thereof) for the needed contract instances.
The output consists of a transaction in the happy case or an appropriate error.
The transaction is then handed over to the \daml{} ledger for validation and state update.

The interpreter loads the referenced \dalf{} packages and their dependencies on demand via the package loader.
The package loader first decodes the Protobuf-encoded \damllf{} code and typechecks it again.
Then, another compilation step further transforms \damllf{} into a simpler non-serializable intermediate language in administrative normal form \cite{FlanaganSabryDubaFelleisen1993PLDI}.

When all required packages have been loaded,
the interpreter converts the commands into $\tyUpdate$ expressions and evaluates them according to the \damllf{} semantics $\evalupd{c}{\_}{\_}$, using an efficient CEK machine implementation~\cite{FelleisenFriedman1987POPL}.
If all evaluations succeed, the overall output consists of the concatenated transactions.

The authorization rules are also checked on the fly while the transaction is built up.
If interpretation tries to exercise a choice, fetch a contract, or create a contract without the required authority,
interpretation fails with an authorization error \emph{before} it starts to evaluate the associated smart contract.
This order matters because otherwise an interpretation error such as a failed \lstinline!assert! or \lstinline!ensure! clause could leak sensitive information via the error message to the submitter.

\subsection{Ledger Integrations}
\label{sec:ledger:integration}

\daml{} smart contracts run on a \daml{} ledger, as shown in \autoref{fig:system:model}.
dApps access the ledger via the Ledger API offered by the participants (as gRPC and HTTP JSON service):
they can submit \daml{} commands and receive synchronous and asynchronous updates of the ledger state.
This event-driven programming model over the uniform Ledger API makes \daml{} dApps portable across different technologies,
as the same \daml{} smart contracts run on all integrations.
To cater for different use cases, integrations with different ledger technologies are available, with different performance, security and confidentiality properties.
We only list the available options for validating transactions and persisting the ledger;
a meaningful comparison is beyond the scope of this paper.
\begin{itemize}
\item A SQL database such as Postgres and Oracle operated by a single trusted entity.
\item A distributed node network run by possibly multiple entities using state machine replication for the deterministic \damllf{} interpreter,
  such as VMware Blockchain \cite{VMwareBlockchain}, Hyperledger Fabric, and Besu.
\item As a second layer on top of a blockchain or consensus algorithm,
  where the contract instances are stored only locally on the participants and the data is encrypted end-to-end when in transit between participants.
  Here, the underlying ledger technology is used only for ordering the updates and coordinating the commits.
\end{itemize}

\section{Related Work}
\label{sec:related:work}

\daml{}'s ledger state is reminiscent of Bitcoin's UTXO model \cite{Nakamoto2008}:
a contract creation corresponds to a ``transaction output'' identified by the contract ID, which can be "spent" by exercising a consuming choice.
In between, a \daml{} contract may be used arbitrarily often via nonconsuming choices, to reduce contention.

Chakravarty et al.\ \cite{ChakravartyChapmanMacKenzieMelkonianPeytonJonesWadler2020FC} proposed the extended UTXO (eUTXO) execution model, which is used by the Plutus \cite{Plutus} smart contract language running on the Cardano blockchain \cite{Cardano}.
In eUTXO, the validation function for spending a transaction output takes an additional input parameter, the datum,
which contains arbitrary contract-specific data.
In \daml{} terminology, this datum corresponds to the contract arguments and the validator function is represented by the choices of the corresponding template.
\daml{} contracts are more flexible in that they are not tied to a cryptocurrency value (the transaction output) and can be used consumingly or nonconsumingly.
As eUTXO transactions are flat, the validators must manually figure out the relevant transaction parts.
\daml{} transactions encode this in the tree structure and thus ensure composability.

Ethereum smart contracts are based on accounts with mutable state rather than UTXO.
Accounts avoid the problem that contract IDs become stale when an update archives the contract instance and re-creates an updated one.
Conversely, accounts suffer from mutability in that the transaction submitter cannot pre-determine the precise effects,
e.g., when the transaction depends on data that changes between submission and execution.
Dangling contract IDs can be avoided in \daml{} by referencing contracts by parts of the stored data instead of by ID (not discussed in this paper).

Corda \cite{Corda} is a DLT platform whose ledger state is a set of immutable contract instances as in \daml{}.
Contract instances are defined using JVM classes that specify both an instance's data and how to validate transactions that create, reference, or consume such an instance.
Corda requires programmers to specify validation as a boolean predicate on whole transactions, which leaves ample room for mistakes:
too lenient checks lead to security vulnerabilities, while too stringent checks hamper workflow composability.
Corda ships contract code as .jar files and executes them on the JVM.
Contract execution on the JVM can be non-deterministic, which may lead to ledger participants getting out of sync; a problem avoided by \daml{}, as it is deterministic by construction.

In the remainder of this section, we compare \daml{} to closely related smart contract languages.
Tolmach et al.\ \cite{TolmachLiLinLiuLi2021Survey} provide a good overview over the wider space.

Peyton Jones et al.\ \cite{PeytonJonesEberSeward2000ICFP} proposed a functional library of small building blocks and combinators to model bilateral financial contracts for valuation purposes.
The contracts of Bahr et al. \cite{BahrBertholdElsman2015ICFP} have a similar focus of valuation and risk analysis, but add the ability to model complex portfolios of multiple parties.
\daml{} picks up this idea of composing simple workflows into more complex ones: choices correspond to alternatives and choice bodies aggregate individual building blocks into larger atomic transactions.

Hvitved et al.\ \cite{Hvitved2011phd, HvitvedKlaedteZalinescu2012JLAP, AndersenBahrHengleinHvitved2014LAFM} developed Peyton Jones et al.'s ideas further into the Contract Specification Language CSL, a process calculus for event-driven transaction systems,
which has been implemented by Deon Digital \cite{DeonDigital}.
CSL stores contract instances as a process term on the ledger, including lambda expressions.
Upon an event, the process term reduces to another term according to the process calculus rules.
This format gives a lot flexibility for combining workflows before instantiation.
In contrast, \daml{} requires the programmer to encode the composition in a new template and produce and deploy the \dalf{} package.
In return, \daml{} contract instances on the ledger are always bound to a template, which helps with understanding the ledger state.
For CSL, it is much harder to figure out what a partially evaluated contract instance represents, as one has to study the raw process term.

While both CSL and \daml{} claim composability, CSL's notion is more restricted:
The CSL composition operators can only be applied before the contract instance is deployed;
building workflows on top of existing contract instances on the ledger is impossible.
So one has to envision all further possible use cases of a contract at the very beginning.
For example, an \lstinline!Iou! can be used in a \lstinline!TradeProposal! workflow only if this was decided when the \lstinline!Iou! contract was created for the first time;
and similarly any follow-up usages.
In contrast, \daml{} allows to compose workflows of deployed contract instances;
the \lstinline!TradeProposal! can work with already deployed \lstinline!Iou!s.
In fact, the \lstinline!Iou! workflows need not even know that they are part of a larger workflow,
thanks to \daml{}'s hierarchical transactions.

In Rainfall \cite{LippmeierRobinsonMuys2019PPDP}, the ledger stores a multiset of tagged facts (template ID and contract arguments in \daml{}) and separately the rules for transforming facts.
Facts are annotated with by-authority and obs-authority, which correspond to \daml{}'s signatories and observers on templates.
A rule pattern-matches on input facts, possibly consuming them, and produces new output facts;
to that end, the rule can acquire by-authority from input facts, similar to how signatories pre-authorize choice consequences in \daml{}.
The use-set stored with each fact determines the rules that may use the fact.
This use-set effectively introduces a dynamic notion of a template:
all facts with the same tag and use-set are instances of the same template.
The main difference between Rainfall and \daml{} is the execution model:
\daml{} references data via unique contract IDs whereas Rainfall finds facts by the content.
So all \lstinline!Iou!s with the same owner, issuer and cash amount are interchangeable in Rainfall.

The Plutus language is based on System F$_\omega$ like \daml{}.
The Plutus compiler from Plutus to Plutus Core uses GHC, too, but with a different setup:
Transaction code is embedded into plain Haskell programs via Template Haskell;
before the GHC backend, a new stage translates these embedded framgents into Plutus Core to be run on-ledger.
The surrounding program is supposed to be run off-ledger.
For \daml{}, we strictly separate the on-ledger code written in \daml{} from the off-ledger code.
To this end, the \daml{} SDK generates bindings for the Java and Typescript, which are used more widely for building off-ledger applications.

\section{Conclusion}
\label{sec:conclusion}

We have presented the smart contract language \daml{} where \lstinline!template!s define the data model for shared on-ledger data together with the \lstinline!choice!s governing how this data may be changed.
The programmer annotates all data and choices with the required authorizers (\lstinline!signatory! and \lstinline!controller!) and the \lstinline!observer!s that get notified about data creation and modification.
These policies not only govern access control, they also partition the data by ownership, which can be exploited for parallel processing.

The \daml{} ecosystem includes more than just the \daml{} language, which this paper focuses on.
The \daml{} SDK includes a VS Code IDE with syntax highlighting, type inference, go-to-definition, etc.
\daml{} Script allows programmers to script scenarios of ledger interactions, as unit tests or for debugging.
\daml{} Triggers provide a simple way to react to ledger updates by sending new commands.
More complex off-ledger logic, including the user interface for the dApp, is typically written in a mainstream programming language and not in \daml{}.
dApps commonly interface with \daml{} code using custom bindings for \daml{} templates and choices, which currently can be auto-generated for Java and TypeScript.

In the future, we plan to extend \daml{} with more flexible ways than explicit \lstinline!DiscloseIou! choices to disclose contract instances to other parties, so that they can exercise their delegated choices.
To improve \daml{}s modularity, we are also working on template interfaces that specify available choices without implementation.
For the \daml{} ledger implementations, we plan to expose some functionality of the underlying blockchain or DLT to the \daml{} code.
In particular, we plan to enable \daml{} smart contracts to directly interact with the native blockchain assets, which is currently only possible via trusted third parties.

\paragraph{Acknowledgements.}

We thank everyone at Digital Asset who contributed to \daml{} and \damllf{}, in particular,
Jost Berthold,
Shayne Fletcher,
Rafael Guglielmetti,
Rohan Jacob-Rao,
Shaul Kfir,
Ben Lippmeier,
Ognjen Mari{\'c},
David Millar-Durrant,
and
Sofus Mortensen.

\bibliographystyle{ACM-Reference-Format}
\bibliography{arXiv}
\end{document}